\documentclass[a4paper]{article}

 \usepackage[pdftex]{graphicx}
\usepackage{amsmath}
\usepackage{wrapfig}

\usepackage{url}

\usepackage{xspace}

\usepackage{enumitem}
\newlist{requirements}{enumerate}{10}
\setlist[requirements]{label=$R\arabic*$}

\newcommand{\streamflow}{StreamFlow\xspace}

\usepackage{authblk}
\title{\streamflow{}: cross-breeding cloud with HPC}
\author[1]{Iacopo Colonnelli}
\author[1]{Barbara Cantalupo}
\author[2]{Ivan Merelli}
\author[1]{Marco Aldinucci}
\affil[1]{Department of Computer Science, University of Torino, Italy}
\affil[2]{Biomedical Technologies (ITB) of the Italian National Research Council
(CNR), Italy}

\date{\scriptsize This paper is the accepted version of IEEE copyrighted material\footnote{
\copyright\ 2020 IEEE. Personal use of this material is permitted. Permission from IEEE must be obtained for all other uses, in any current or future media, including reprinting/republishing this material for advertising or promotional purposes, creating new collective works, for resale or redistribution to servers or lists, or reuse of any copyrighted component of this work in other works.}\\
I. Colonnelli, B. Cantalupo, I. Merelli, and M. Aldinucci, “Streamflow: cross-breeding cloud with HPC,” IEEE Transactions on Emerging Topics in Computing, 2020. DOI:10.1109/TETC.2020.3019202.}                     
\begin{document}

  \maketitle

\begin{abstract}
  Workflows are among the most commonly used tools in a variety of execution environments. Many of them target a specific environment; few of them make it possible to execute an \emph{entire} workflow in different environments, e.g. Kubernetes and batch clusters. We present a novel approach to workflow execution, called \streamflow{}, that complements the workflow graph with the declarative description of potentially complex execution environments, and that makes it possible the execution onto multiple sites not sharing a common data space. \streamflow{} is then exemplified on a novel bioinformatics pipeline for single-cell transcriptomic data analysis workflow.
  \end{abstract}
  
  \section{Introduction}
  
  {Both} in the HPC and cloud realms, workflows play an essential role for applications coordination because they provide means to model and formalise complex processes in multiple steps, e.g. tasks, jobs, OS containers or even Virtual Machines, depending on the target system. Steps are generally arranged in a partial order induced by (true) data dependency. For this, workflows can be naturally represented with direct graphs.   
  
  Although workflows are used in different execution environments, such as HPC, cloud and edge, all of these environments continue their path toward greater specialisation in term of typical features and workloads. While RESTful APIs are becoming the lingua franca to access and compose computation and storage in the cloud, the HPC platforms are bound to batch job schedulers. Starting a web server on an HPC platform is generally not admitted, as it is impractical to access to cloud storage, e.g. to retrieve temporary results.  While the execution of \emph{independent} steps in the cloud means they can be executed \emph{in any temporal order in a single processing element}, in the HPC platforms the need for co-allocating \emph{at the same time} multiple processing elements to execute a single job is the rule \cite{stkm:europar:09}. This complementarity is the cornerstone of a computing continuum that appears emerging in data-driven applicative domains. We envision this continuum as composed of more and more specialised and therefore heterogeneous environments. For this, also workflows need to embrace heterogeneity, by embedding the capability to execute a single workflow on multiple different environments. For this to happen, workflows should gain a higher level of abstraction, subsuming the role of coordination language of other lower level and more specialised workflows targeting a specific platform.     
  
  In this work, we introduce \emph{\streamflow{}}, a novel workflow model that extends a classic workflow system with a declarative description of possibly many environments and with the relations among workflow nodes and execution environments. \streamflow{} is not yet another workflow system; it somewhat conceptually aims at complementing a workflow system to raise its level of abstraction, providing the workflow with a ``virtual'' cross-site platform.  In other words, \streamflow{} makes it possible to partition a workflow and describe an execution plan spawning across multiple sites, even if they do not share the same data space. In this, \streamflow{} leverages on lower level features such as the deployment of explicitly parallel nodes, e.g. MPI execution, which is targeted via HPC jobs schedulers (supporting OS containers).
  
  The \streamflow{} concept is exemplified by way of a proof-of-concept implementation based on the Common Workflow Language (CWL) interface, which is used to specify a novel bioinformatic pipeline (single-cell transcriptomic data analysis). Thanks to \streamflow{}, the single-cell pipeline is executed on two sites: a Kubernetes orchestrator on the cloud and an HPC cluster on-premise.    
  
  In Sec.~\ref{sec:rel} we describe related work. Being the literature in workflows massive, we focus on the aspects of interest for this work, inviting the reader to refer to existing surveys for a more general comparison among workflow systems. In Sec.~\ref{sec:methods}, we present the proposed approach, i.e. \streamflow{} basic principles, whereas \streamflow{} design and implementation are described in Sec.~\ref{sec:impl}. Sec.~\ref{sec:usecase} reports the single-cell transcriptomic data analysis workflow, along with \streamflow{} experimentation. Finally, Sec.~\ref{sec:conclusion} summarises conclusions and future works.
  
  
  \begin{section}{Related works}
  \label{sec:rel}

    Workflows provide powerful abstractions to design scientific applications, also supporting their execution on specific infrastructures. According to this vision, we can consider workflows as an interface between the domain specialists and the computing infrastructure. The Workflow Management System (WMS) landscape is very variegated, as it embraces scientific domain tools, mainly focused on resolving typical modelling issues in the domain, and low-level specifications, aimed at executing tasks on multi processes infrastructures. Several surveys exist on WMSs, comparing their different functionalities \cite{Cohen-Boulakia:2017,GridSurvey:2015}, focusing  on their evolution \cite{Atkinson:2017} or providing classification with respect to the support for extreme-scale applications \cite {FERREIRADASILVA:2017228}. In the context of this work, we are particularly interested in understanding the most critical needs, the most effective approaches and the most promising developments in this continuously changing technological domain.
    
    In particular, two main levels of analysis should be considered: the application level, where the orchestration of the different functional components of the application is managed, and the infrastructure level, where the computational units composing the workflow are executed by the workflow engine. At the first level, it is essential to evaluate the ability of the system to respond to user needs, by supporting potentially complex multi-node execution environments, and manage massive amounts of data ingested and computed by all the applications. At the infrastructure level, together with established architectures like clusters or grids, the cloud is now the most referred infrastructure for application execution and new paradigms are gaining attention like containers and orchestrators. Finally, HPC facilities are  getting more and more importance outside the research centre even if there is no straight road-map for their integration with other platforms till now. 
    
    \begin{subsection}{Scientific workflows} \label{subsec:scientific-workflows}
      WMSs for scientific workflows are user-driven systems specifically developed to satisfy domain requirements. They provide researchers with a useful paradigm to describe, manage and share complex scientific analyses, also ensuring reproducibility and scalability properties. Experiments can be modelled by using a high-level declarative language or advanced graphical interfaces, suitable for researchers with little programming experience, or described programmatically.
      
      Many scientific WMSs (e.g. Kepler\footnote{\url{https://kepler-project.org/}} \cite{Kepler:2006}, Askalon\footnote{\url{http://www.askalon.org/}} \cite{Askalon:2007},  Pegasus\footnote{\url{https://pegasus.isi.edu/}} \cite{pegasus:15},  Taverna\footnote{\url{https://taverna.incubator.apache.org/}} \cite{taverna:06} and Galaxy\footnote{\url{https://galaxyproject.org/learn/advanced-workflow/}} \cite{galaxy:16}) emerged with the diffusion of the web services and grid technologies, which offered the possibility to access robust services and infrastructures in a more natural way than before \cite{Badia:2017}. Therefore, they were mainly targeted towards these architectures and not focused on portability. Nevertheless, by evolving in strict contact with the scientific community, they acquired maturity from the functional design point of view and started providing some additional features, as workflow repositories or support for diverse newer architectures, establishing consensus among researchers.
      
      Even if some of these tools like Pegasus and Askalon offer support for automatic data transfers also in the absence of a commonly shared file-system among worker nodes, they rely on specific transfer protocols (e.g. GridFTP, SRM or Amazon S3) or delegate it to an external batch scheduler such as HTCondor, actually constraining the set of supported configurations. Moreover, although both Galaxy and Pegasus offer support for container execution, they only allow mapping a task into a single container. Asterism \cite{Asterism:2019}, a hybrid framework where the stream-based workflow execution is managed by dispel4py \cite{dispel4py:2017} and the data movement among workers is left to Pegasus, represents an interesting exception. Indeed, it relies on Docker Compose\footnote{\url{https://docs.docker.com/compose/}} to set up complex execution clusters. Nevertheless, at the time of writing, dispel4py only provides MPI and Apache Storm\footnote{http://storm.apache.org/} executors, strongly limiting the potential of the library.
      
      An alternative approach to complex and feature-rich WMSs privileges performances over accessibility, exposing lower-level programming models directly to the user and allowing for the execution of a large number of interconnected tasks on distributed architectures. Among these frameworks are HyperLoom \cite{hyperloom:18} and Dask\footnote{\url{https://docs.dask.org/}}  \cite{dask:2015}, where pipelines of tasks can be defined with a Python interface, Spotify's Luigi\footnote{\url{https://github.com/spotify/luigi/}}, which also comes with a visual interface for monitoring purposes, and COMP Superscalar (COMPSs) \cite{compss:12}, that allow users to parallelise existing sequential applications by identifying and annotating functions that can be executed as asynchronous parallel tasks. Despite being very efficient in terms of performances, these libraries are hard to use for domain experts without programming experience.
      
      Another approach tries to lower the level of complexity by implementing a simplified Domain Specific Language (DSL) to describe workflows. For example, in Apache Airflow\footnote{\url{https://airflow.apache.org/}} and  Snakemake\footnote{\url{https://snakemake.readthedocs.io/en/stable/}} \cite{snakemake:12} workflows are essentially Python scripts extended by declarative code that can be executed on distributed infrastructures. Other systems adopt Unix-style approaches for defining workflows: in Makeflow\footnote{\url{http://ccl.cse.nd.edu/software/makeflow/}} \cite{makeflow:12} the end-user expresses a workflow in a technology-neutral way using a syntax similar to Make, while the Nextflow\footnote{\url{https://www.nextflow.io/}} \cite{nextflow:17} bioinformatics framework builds workflows on Unix pipe concept. In these frameworks, a set of pluggable executors allows workflows to be deployed and run on different infrastructures, including public cloud services, batch schedulers (e.g. HTCondor, PBS, SLURM) and Kubernetes clusters. Nevertheless, different steps of the same workflow cannot be managed by different executors, not guaranteeing support for hybrid cloud/HPC configurations. Moreover, even if containers executions are permitted, it is not possible to specify complex multi-container environments to execute a single task.
      
      Since product-specific DSLs tightly couple workflow descriptions to a single software, actually limiting portability and reusability, there are also efforts in defining workflow specification languages or standards. For instance, The Common Workflow Language (CWL)\footnote{\url{https://www.commonwl.org}}  \cite{cwl:1} is an open standard for describing analysis workflows following a JSON or YAML syntax or a mixture of the two. One of the first and most used CWL implementations is CWL-Airflow \cite{cwl-airflow:19}, which adds support for CWL to Apache Airflow, but also other products (e.g. Snakemake and Nextflow) offer some compatibility with CWL. Another interesting dataflow language for scientific computing is Swift\footnote{\url{http://swift-lang.org/main/}} \cite{swift:2014}, in which all statements are eligible to run concurrently, limited only by the data flow.

    \end{subsection}
    
    \begin{subsection}{Cloud orchestration} \label{subsec:cloud-orchestration}
      With the rise of cloud computing and the related *-as-a-Service approaches, users from both academia and industry started to move computation from their machines to the cloud. Nevertheless, all the scalability, portability and availability benefits brought by cloud technology necessarily come with increased complexity in deploying, configuring and managing applications, especially in hybrid-cloud scenarios.
      
      Several tools have been developed with the precise aim to seamlessly support hybrid-cloud deployments of complex applications, composed of heterogeneous intercommunicating services, by exposing to the user just an agnostic and straightforward interface, usually in the form of a DSL. This approach is often referred to as \textit{Infrastructure-as-Code} (IaC).
      
      Early solutions \cite{Chieu:2010,Lenk:2011} focused their attention on the deployment phase, with minimal support for contextualisation (a simple run-once script launched right after the virtual machine creation) and no support at all for automatic orchestration of active applications. In this setting, both the creation and maintenance of ad-hoc base images for each required software application and all the post-deployment operations were left to the IT experts.
      
      More recently, some more advanced tools have been proposed, addressing different aspects of application deployment and orchestration. Caballer et al. \cite{Caballer:2015} focus on reusability, introducing a platform in which every configuration element, from infrastructure descriptions to virtual machine images, can be stored in a dedicated registry. SALSA \cite{Le:2014} comes with a fine-grained, multi-layered dependency structure, in order to handle the configuration of both virtual machines and applications deployed upon them. Roboconf \cite{Pham:2015} proposes a hierarchical DSL, capable of defining both containment and runtime relations, and offers orchestration primitives to automatically manage the reconfiguration of live systems in response to events, e.g. an increased workload or a failure. Occopus \cite{Kovacs:2018} privileges compatibility and extendibility, proposing a pluggable architecture in which combinations of replaceable plugins manage interactions with external tools and services (as well as some core features and behaviours). Moreover, it comes with some orchestration features, e.g. health-checking, auto-scaling and garbage collection. 
      
      Workflow management and cloud orchestration technologies can benefit from each other. For example, the OASIS Topology and Orchestration Specification for Cloud Applications (TOSCA)\footnote{\url{https://www.oasis-open.org/}} \cite{TOSCA:2013} focuses on complex dependency management, using workflow description languages to write the deployment and management plans for a cloud environment. Implementations of the TOSCA standard are currently provided by Cloudify\footnote{https://cloudify.co/}, an orchestration platform based on event-driven workflows, and Yorc\footnote{https://github.com/ystia/yorc} (Ystia orchestrator), an hybrid cloud/HPC orchestrator developed in the Lexis project \cite{Lexis:2020}. Moreover, a Cloudify plugin to orchestrate batch applications in HPC and cloud environments has been developed in the Croupier\footnote{https://github.com/ari-apc-lab/croupier} project.
      
      With the advent of containerisation as a lightweight alternative to virtualisation, some container orchestrators started to flourish. Among them, Kubernetes has become the de-facto standard for container orchestration during the last years, and the vast majority of cloud providers include a managed Kubernetes service in their offering. Kubernetes comes with a very flexible YAML-based DSL, able to describe both deployment and runtime orchestration features for multi-container applications. Containers are also gaining popularity  in the scientific domain, and several workflow frameworks have been built natively on top of Kubernetes like Pachyderm\footnote{\url{http://pachyderm.io/​}} \cite{pachyderm:18}, Argo\footnote{\url{https://argoproj.github.io/argo/}} and a specific Galaxy installation developed in the PhenoMeNal\footnote{\url{http://phenomenal-h2020.eu/home/​}} project \cite{galaxy-kub:18}. Moreover, all the leading cloud vendors are currently focusing on offering hybrid solutions that allow combining multi-cloud and on-premises infrastructures. The most integrated framework is  GoogleCloudComposer\footnote{https://cloud.google.com/composer}, a fully managed workflow orchestration service built on Apache Airflow. Despite offering great flexibility in interacting with Kubernetes resources, these products are tightly coupled with such technology and do not allow for task offloading on different environments, such as HPC sites.
      
      Even if both WMSs and orchestrators must be able to deal with dependencies among different tasks, they differ in their primary goals. Indeed, from one side, WMSs must focus on efficient ephemeral executions of tasks, minimisation of the distributed execution overheads (e.g. through data-locality-based scheduling policies) and should be easy enough to be used by domain experts. From the other one, orchestrators must ensure availability and responsiveness of long-lived systems, portability among different infrastructures and enough flexibility to satisfy the needs of IT experts. With this in mind, StreamFlow aims to offer an easy way to allow the automatic execution of workflows on top of complex and orchestrated environments while keeping the two aspects distinct enough to be easily handled by different kinds of users.
    \end{subsection}
  \end{section}
  
  \begin{section}{Methods}
  \label{sec:methods}
    \begin{subsection}{Multi-container environments} \label{subsec:multi-container-environments}
      Portability and reproducibility have always been two fundamental aspects of scientific workflows. Nevertheless, the combination of the two is undoubtedly a non-trivial requirement to satisfy, since it is necessary to guarantee that a piece of code running on top of potentially very diverse execution environments will give identical results. The first obvious issue here comes from the need to provide the same versions of all the libraries directly or indirectly involved in the computation. On top of that, some numerical stability problems can arise when running the same code on different platforms, e.g. on Linux and Mac OS X \cite{nextflow:17}. Fortunately, with the diffusion of lightweight containerisation technologies like Docker\cite{Merkel:2014} and Singularity\cite{Kurtzer:2017}, a straightforward solution for these issues finally appeared and nowadays container-based tasks are supported by a wide number of WMSs on the market, either as an alternative to native execution or as first-class citizens \cite{Kulkarni:2018}.
  
      The typical way to support containerisation in WMSs is through a one-to-one mapping between tasks and containers, i.e. a container image is associated with each task in the workflow graph. In this setting, the execution flow of a single task always consists of three sequential steps: the container is launched, the task is executed inside it, and finally, the container is stopped. Drawing a parallel with the famous Flynn's taxonomy \cite{Flynn:1972}, we could define this execution pattern as \textit{Single-Task Single-Container} (STSC).
      
      When compared with a \textit{Multiple-Tasks Single-Container} (MTSC) alternative, the STSC pattern comes with a decisive advantage. Since containers' file-system is commonly ephemeral, every task execution runs inside a clean and consistent environment (with the apparent exception of eventual temporary files saved into persistent folders). For its part, an MTSC execution can provide some performance improvements in those cases when the task execution is high-speed (comparable with the startup and shutdown overheads of a container, generally in the order of milliseconds). Moreover, MTSC can be useful also when a process inside the container must complete a heavy initialisation phase before being ready to perform tasks or when some data dependencies are stored in the ephemeral file-system, in order to avoid additional data transfers when recreating containers.
      
      Far more interesting would be the \textit{Single-Task Multiple-Containers} (STMC) setting, because it allows using multiple, possibly heterogeneous environments to solve a single task. For example, with an STMC approach, it would be possible to run an MPI task on top of multiple nodes or a MapReduce-based task with multiple instances of Apache Spark.
      
      Finally, the most general setting of \textit{Multiple-Tasks Multiple-Containers} (MTMC) would also allow for \textit{concurrent} task execution, i.e. a configuration in which tasks $T_1$ and $T_2$ execute at the same time on different resources and $T_{1}$ produces data consumed by $T_2$. The support for this last configuration becomes fundamental when dealing with stream-based workflows \cite{FERREIRADASILVA:2017228}. In principle, also an MTSC configuration enables the concurrent execution of tasks into the same resource, but here the advantage is less valuable. Indeed, it is far easier to obtain the same behaviour in an STSC setting with a single task charged with launching and managing all the required processes.
  
      Unfortunately, a simple many-to-many task-image association is not enough to model an *MC configuration, because it is also necessary to explicitly specify the connections among different containers. Nevertheless, some ways to define multi-container environments are already present on the market, from simple libraries like Docker Compose and Singularity Compose\footnote{\url{https://singularityhub.github.io/singularity-compose/}} to complex orchestrators as Kubernetes\footnote{\url{https://kubernetes.io/}} or Docker Swarm\footnote{\url{https://docs.docker.com/engine/swarm/}}. Therefore, it is a wise choice to rely on them for the environment definition. This can be achieved by substituting the original one-to-one task-container association with a many-to-one task-environment association and by treating an entire multi-container environment as the unit of deployment. It is worth noting that even a many-to-many association would be potentially feasible, allowing to split a single task among different environments. Nevertheless, this would overcomplicate both the scheduling policies and the communication layer, forcing the need to distinguish between inter-environment and intra-environment interactions among different resources executing the same task.
  
      The following two requirements can summarise all these considerations:
      \begin{requirements}
        \item \label{req:deployment-unit} A uniquely identified multi-container environment definition must be treated as an atomic deployment unit. A unit must be deployed before starting to execute the first associated task and undeployed after the execution of the last associated task.
        \item Each task can be associated with a single deployment unit, but the same deployment unit can be associated with multiple tasks.
      \end{requirements}
    \end{subsection}
  
    \begin{subsection}{Hybrid workflows} \label{subsec:hybrid-workflows}
      When considering data-intensive scientific workflows, all those aspects related to data management (such as data locality, data access, and data transfers) become crucial as well. In this setting, the need for a WMS capable of dealing with \textit{hybrid workflows}, i.e. to coordinate tasks running on different execution environments \cite{FERREIRADASILVA:2017228}, can be a crucial aspect for performance optimisation when working with massive amounts of input data. Indeed, an \textit{in situ} data processing strategy can prevent all the overheads related to data transfers and even to disk I/O when in-memory processing is allowed. Moreover, hybrid workflow execution becomes a mandatory requirement when dealing with federated data access or strict privacy policies.
  
      Even if many of the existing WMSs can run the same workflow with a diverse set of \textit{executors}, some of them addressing cloud environments and some others more HPC-oriented, a far smaller percentage of them can deal with multi-cloud and hybrid cloud/HPC execution environments for a single workflow. The first step to take in this direction is to waive the requirement for any shared data access abstraction among all the containers, keeping the only constraint for the WMS management node to be able to reach the whole execution environment. Such a scenario provides a significant amount of flexibility. Unfortunately, it implies that every inter-container data transfer needs at least two copy operations: a first one from the source to the management node and a second one to the destination. Sometimes this is the only way to go, but if direct communications between container pairs are possible, then it could be better to rely on them if only to avoid overloads on the central management node. Therefore, the best strategy here would probably be to consider the two-steps copy proposed above as a baseline communication channel between every container pair while allowing users to declare better ways to exchange information when available.
  
      From a practical point of view, the logic related to data transfers can be specified at two different levels:
      \begin{itemize}
        \item At the \textit{host language level}, i.e. directly embedded in the business logic of the producer task. In this scenario, the only thing that the WMS can do is to check for the existence of the expected destination path before starting the data transfer process, in order to avoid useless overheads.
        \item At the \textit{coordination language level}, i.e. explicitly specified by the user in the workflow description. In this scenario, the management of data transfers is left to the WMS, which can rely on a dedicated channel or fall back to the baseline strategy, as discussed above. 
      \end{itemize}
      While the former case is quite easy to implement, the latter would require a channel abstraction, flexible enough to manage different data types (from simple values to huge file-system portions) and to deal with the aforementioned multi-container environments, potentially deployed on multi-cloud or hybrid cloud/HPC architectures. For now, to keep things a bit simpler, we decided to always rely on the baseline strategy for the inter-environment case, while implementing slightly more optimised solutions for the intra-environment case whenever possible. Nevertheless, a better language specification for communication channels is, for sure, one of the most critical future improvements for the proposed approach.
  
      Again, the following two requirements can summarise the previous discussion:
      \begin{requirements}[start=3]
        \item If the WMS management node can reach the whole execution environment, then an inter-container data transfer must always be possible, with a two-steps copy operation as the baseline strategy. Optimisations are possible for intra-environment data transfers.
        \item If data are already present in the destination path, the WMS should avoid performing an additional copy.
      \end{requirements}
    \end{subsection}
  \end{section}
  
  \begin{section}{\streamflow{} framework}
  \label{sec:impl}
    \begin{figure}[t]
      \begin{center}
        \includegraphics[width=.9\textwidth]{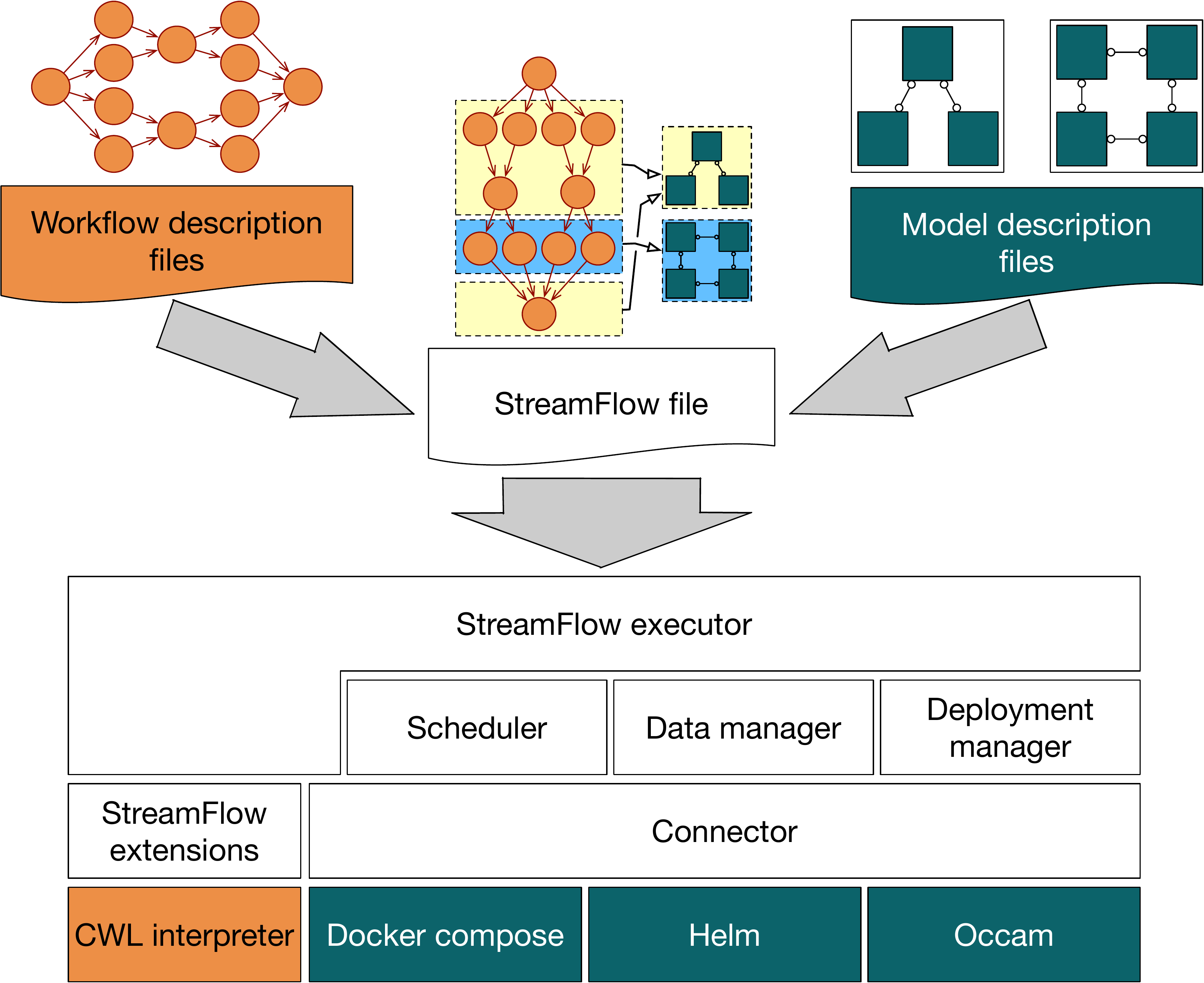}
        \caption{\streamflow{} framework's logical stack. Coloured portions refer to existing technologies, while white ones are directly part of \streamflow{} codebase. In particular, the orange area is related to the definition of the workflow's dependency graph, while the green area refers to the execution environments.}
        \label{fig:streamflow-model}
      \end{center}
    \end{figure}
    The \streamflow{} framework\footnote{\url{https://streamflow.di.unito.it/}} has been created as a proof-of-concept WMS based on the four previously discussed requirements. Written in Python 3, it has been designed to seamlessly integrate with existing WMSs' coordination languages, in order to allow users to extend their existing workflows without having to change what has been already done. In keeping with this point of view, we also decided not to define a new description language for multi-container environments, but rather to build a common interface to allow for the integration with existing technologies.
    
    In \streamflow{}'s glossary, a complex multi-container environment is called \textit{model}. Each model is managed independently of the others by a dedicated \texttt{Connector} implementation, which acts as a proxy for the underlying orchestration library. A single model can include multiple types of containers, called \textit{services}. For example, a Docker Compose file describing a database and a Tomcat container linked together constitutes a model with two services. The \texttt{streamflow.yml} file, the actual entry point for a \streamflow{} execution, contains pointers to workflows and models descriptions and specifies the way they should relate to each other, i.e. the service that should execute each workflow \textit{step}. Since  multiple replicas of the same service could coexist in a given model, each service can refer to one or more containers, called \textit{resources}.

    Before actually executing a task, it is necessary to deploy the related model successfully. The \texttt{DeploymentManager} class has precisely the role of creating models when needed and destroying them as soon as they become useless. Then the \texttt{Scheduler} class is in charge to select the best resource on which each task should be executed while guaranteeing that all requirements are satisfied. Finally, the \texttt{DataManager} class, which knows where each task's input and output data reside, must ensure that each service can access to all the data dependencies required to complete the assigned task, performing data transfers only when necessary. At this point, a \textit{job} (i.e. the runtime representation of a task) can be successfully executed on the selected resource.
    
    The rest of the current section is devoted to analysing with more detail each of the components mentioned above, whose position in the \streamflow{}'s logical stack is represented in Fig. \ref{fig:streamflow-model}, and how they coordinate with each other.
  
    \begin{subsection}{The WMS integration layer} \label{subsec:wms-integration-layer}
      As stated before, one of the design choices for the \streamflow{} approach is to rely on existing coordination languages, instead of coming with yet another way to describe workflow models. In order to realise a first proof-of-concept, we decided to integrate with the CWL format. Being a fully declarative language, CWL is far simpler to understand than its Make-like or dataflow-oriented alternatives. Moreover, some existing WMSs provide at least a partial compatibility with CWL format, even when it is not their primary coordination language. Last but not least, the CWL's reference implementation, called cwltool\footnote{\url{https://github.com/common-workflow-language/cwltool}}, is written in Python: this not only allowed us to use the official library to obtain the compiled workflow representation, but also to rely on existing classes for the main part of the execution process.
      
      Therefore, what we did in practice was to provide an extension layer to the original cwltool codebase, using inheritance to inject additional features or to override the existing ones whenever required. This approach considerably reduced the development time, but the risk is to introduce excessively tight coupling between the CWL-specific features and the more generic \streamflow{} logic. Since we plan to support other coordination languages in the future, a more agnostic mid-layer representation of a workflow graph is definitely on the todo list.
    \end{subsection}
 
    \begin{subsection}{Model life-cycle management} \label{subsec:model-life-cycle-management}
      In \streamflow{}, the service allocation and the subsequent task execution happen in two strictly distinct phases, leaving the containers' life-cycle management to an external orchestration library. A clear advantage of this approach lies in the possibility to rely on all the orchestration features provided by a mature product (e.g. autoscaling, restarting policies, affinity-based scheduling) and to adopt the original deployment description language, sparing users the extra effort needed to learn a new syntax.  Moreover, as behind the scenes \streamflow{} demands the deployment and undeployment phases to the original orchestrator, there are no constraints on the supported features: if it works with the original library, it works with \streamflow{}.

      \begin{figure}[t]
        \begin{center}
          \includegraphics[width=.9\textwidth]{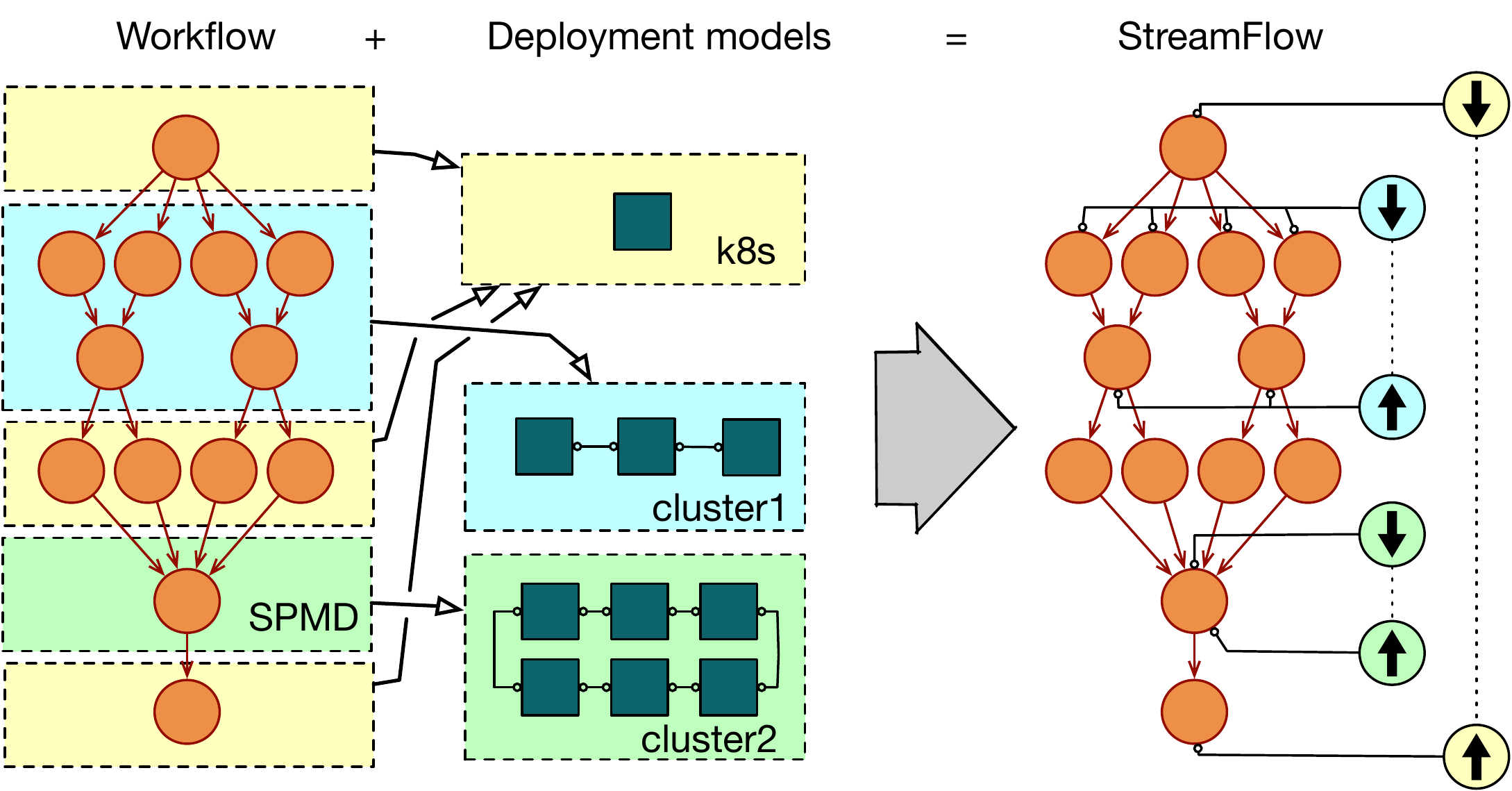}
          \caption{Workflow graph transformation to include model deployment and undeployment tasks. Orange nodes represent original tasks, while the others refer to model deployment (downward pointing arrow) and undeployment (upward pointing arrow) phases.} 
          \label{fig:streamflow-graph-transformation}
        \end{center} 
      \end{figure}      
      As shown in Fig. \ref{fig:streamflow-graph-transformation}, from a theoretical point of view, this approach can still be represented with a traditional workflow model, by transforming the original dependency graph in order to include two new special kinds of tasks:
      \begin{itemize}
        \item The \textit{deployment} task, which synchronously creates a new model. This task does not depend on anything else, but all the tasks that should be executed in such a model must depend on it.
        \item The \textit{undeployment} task, which destroys an existing model. No other task depends on it, but it should depend on all the tasks that must be executed on such a model, in order to wait for their termination before starting the undeployment process.
      \end{itemize}
      The result of this transformation is a perfectly fine dependency DAG, which satisfies requirement $R1$ and can be correctly described by the vast majority of coordination languages. Nevertheless, since deployment tasks have no dependency, a standard scheduler will try to execute them as soon as possible, according to an \textit{eager} resource allocation strategy. In this setting, some models can be up and running long before they are needed, leading to a potential waste in terms of energy consumption and money. In such case, a far more practical approach would be to let a model be deployed by the first fireable task which requires it, according to a \textit{lazy} resource allocation strategy.
  
      \begin{figure}[t]
        \begin{center}
          \includegraphics[scale=0.60]{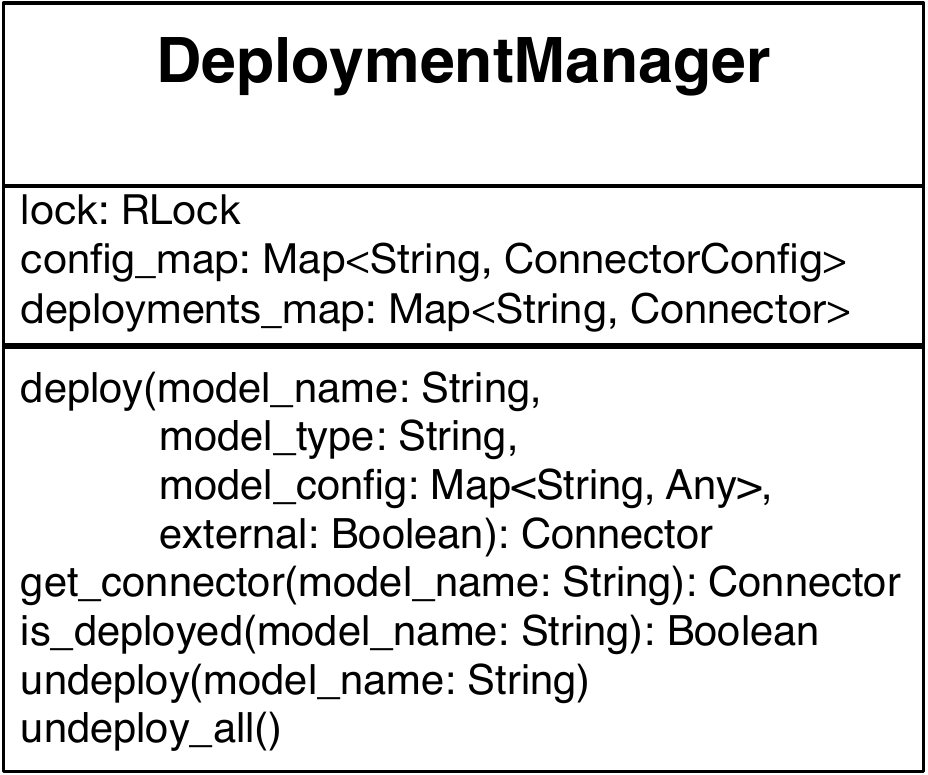}
          \caption{UML class diagram for the \texttt{DeploymentManager} class.} 
          \label{fig:deployment-manager-uml-model}
        \end{center} 
      \end{figure}
      The \texttt{DeploymentManager} class, whose UML diagram is represented in Fig. \ref{fig:deployment-manager-uml-model}, has precisely the role of implementing these allocation strategies, relying on the underlying orchestration library through a pluggable implementation of the \texttt{Connector} interface. In particular, the \texttt{deploy} method atomically checks if a model has been already deployed and, if not, it puts a new \texttt{Connector} instance into the \texttt{deployments\_map} and invokes its \texttt{deploy} method. Since the requirement $R2$ states that a single instance of a model can be used to execute multiple tasks, the \texttt{lock} is necessary to avoid race conditions when concurrent tasks require the same model. Finally, the \texttt{external} attribute allows \streamflow{} to interact with an externally managed model, relieving the \texttt{DeploymentManager} of deployment and undeployment duties.
      
      Ideally, a model should be undeployed as soon as the last task needing it has been completed. This logic is quite easy to implement when dealing with static DAGs, but things get more complicated in the dynamic setting. Probably the best strategy for the second case would be to set a grace period, after which the model is undeployed if no new task required it. For now, the \texttt{DeploymentManager} confines itself to undeploy all the models at the end of the entire workflow execution, calling the \texttt{undeploy\_all} method. Moreover, the same method is also invoked by \streamflow{}'s main process in case of unrecoverable failures. This approach is very straightforward, but it can lead to resource wastes if some models remain unused for a long time.
      
      \begin{figure}[t]
        \begin{center}
          \includegraphics[scale=0.60]{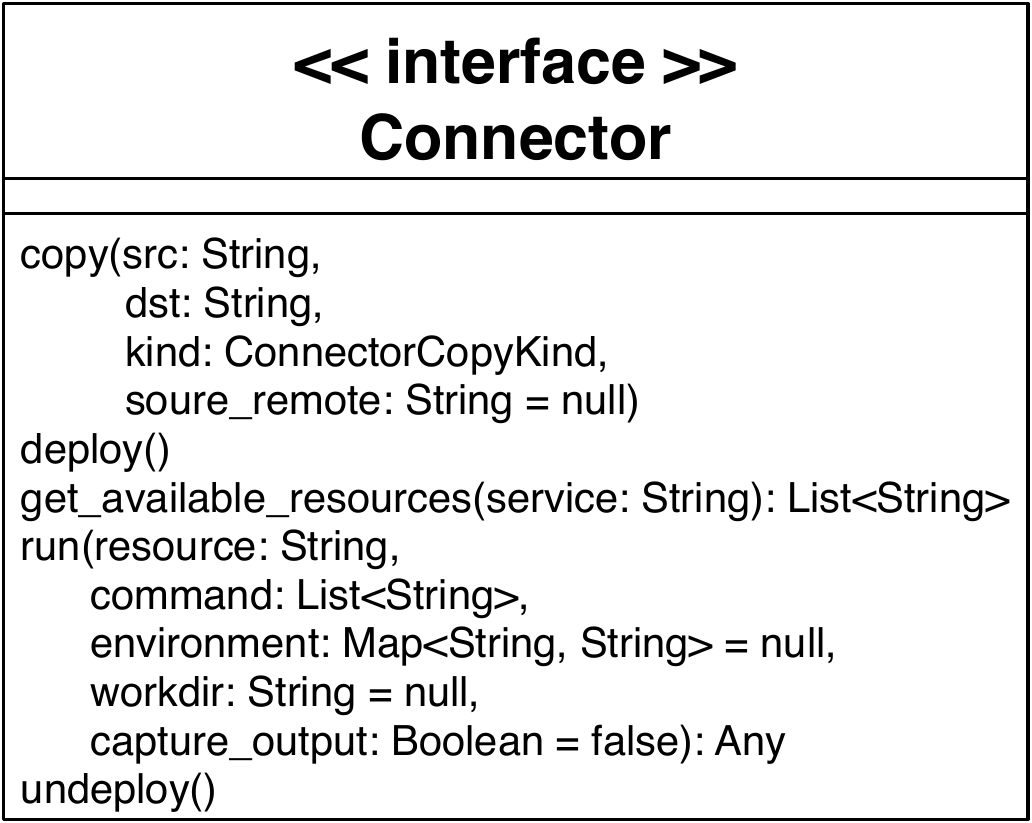}
          \caption{UML class diagram for the \texttt{Connector} interface.} 
          \label{fig:connector-uml-model}
        \end{center} 
      \end{figure}
      As discussed before, \streamflow{} interacts with each underlying orchestration technology by means of a common \texttt{Connector} interface, whose UML diagram is shown in Fig. \ref{fig:connector-uml-model}. This adheres to the separation of concerns principle, providing an easy way to add support for additional products if required. The \texttt{Connector} interface is a low-level block in the \streamflow{}'s logical stack, which is used by all the higher-level components. Besides the \texttt{deploy} and \texttt{undeploy} methods, which are called by the \texttt{DeploymentManager} class, the \texttt{get\_available\_resources} method is invoked by the \texttt{Scheduler} class to obtain all the replicas of a given service in the model, while the \texttt{copy} method is instead used by the \texttt{DataManager} class to perform data transfers among resources, with the \texttt{kind} argument specifying the direction of the transfer operation. Finally, the \texttt{run} method is used to execute a command on top of a remote resource and potentially to capture the generated output value. For now, three different \texttt{Connector} implementations come out-of-the-box with \streamflow{}, supporting Docker Compose, Helm\footnote{\url{https://helm.sh/}} and Occam, the supercomputing centre of Universit\`a di Torino \cite{16:occam:chep}.
    \end{subsection}
  
    \begin{subsection}{The \streamflow{} file}\label{subsec:streamflow-file}
      When launching a \streamflow{} execution, the only argument it takes is the path of a YAML file, conventionally called \texttt{streamflow.yml}. The crucial role of such file is to link each task in a workflow with the service that should execute it. Moreover, in order to ensure this binding is unambiguous, each service in a model and each task in a workflow should be uniquely identifiable. This section describes the \streamflow{} file syntax and the strategies adopted to guarantee such unambiguity.
      
      A valid \streamflow{} file contains the version number (which currently only accepts the \texttt{v1.0} value) and two main sections: \texttt{workflows} and \texttt{models}. The \texttt{workflows} section consists of a dictionary with uniquely named workflows to be executed in the current run. Each workflow specification is an object containing three fields. The \texttt{type} field identifies which language has been used to describe the dependency graph (at the moment \texttt{cwl} is the only accepted value), while the \texttt{config} field includes the paths to the files containing such description. Finally, the \texttt{bindings} list contains the task-model associations. Different workflows are independent of each other, in that an entire \streamflow{} logical stack is allocated for each of them. It means that, even if two tasks in two different workflows can refer to the same model specification, two different environments will be deployed for their execution.
      
      Considering workflows as dependency graphs, each node can refer to either a simple task or a nested sub-workflow. Therefore, we decided to adopt a file-system based mapping of each task to a Posix-like path, where each simple task is mapped to a file, and each sub-workflow is mapped to a folder, which can contain both files and sub-folders. In particular, the most external workflow description is mapped to the root folder. Such method allows for easy and unambiguous identification of tasks, given that there exists an intuitive way to assign a name to each task in the workflow's graphical structure and that such name has the uniqueness constraints required by a typical file-system representation. Fortunately, CWL standard (and also the vast majority of coordination languages on the market) satisfies both these requirements. 
      
      The \texttt{models} section contains a dictionary of uniquely named model specifications, each of which is an object with two distinct fields. The \texttt{type} field identifies which \texttt{Connector} implementation should be used for its creation, destruction and management, while the \texttt{config} field contains a dictionary with configuration parameters for the corresponding \texttt{Connector}. Usually, the \texttt{config} parameters are directly extracted from the tools commonly used to interact with the underlying orchestration library (e.g. the \texttt{docker-compose} CLI for Docker Compose or the \texttt{helm} CLI for Helm charts), so that a user who is familiar with these libraries can easily understand the \streamflow{} format.
      
      The best way to unambiguously identify services in a model strictly depends on the model specification itself. For Docker Compose, where the unit of deployment is a single container, it is enough to take a key in the \texttt{services} dictionary to identify the related service uniquely. Moreover, since an Occam description file is practically equivalent to the \texttt{services} section of a Docker Compose file, the same strategy can be applied to it, too. Unfortunately, in Kubernetes (and consequently in Helm) the unit of deployment is a Pod, which can contain multiple containers inside it. In this case, the user is explicitly required to fill in the \texttt{name} attribute of each container in the Pod template with a unique identifier.
  
      The format adopted for the \texttt{bindings} list takes into account all the previously discussed considerations on unambiguous identification of tasks and services. In particular, each element of such list contains a \texttt{target} object, with a \texttt{model} and a \texttt{service} attributes that uniquely identify a service, and a \texttt{step} attribute containing a path in the aforementioned file-system abstraction of a workflow graph. If the path resolves to a folder (i.e. to a nested sub-workflow), the same target service is applied recursively in the file-system hierarchy, unless a more specific configuration (i.e. another entry in the \texttt{bindings} list with a deeper path in its \texttt{step} field) overrides it. For the interested reader, the whole specification for the current version of the \streamflow{} file is contained in a JSON Schema file named \texttt{config\_schema.json}. Since such file is also used in the validation phase during a \streamflow{} execution, it represents the authoritative source of truth for the \streamflow{} file format.
    \end{subsection}
  
    \begin{subsection}{Task scheduling}\label{subsec:task-scheduling}
      The task scheduling strategy is a fundamental component of a WMS, mainly for the large impact it has on the overall execution performances. It is a common practice for WMSs to allow users to specify some minimum hardware requirements for a task, e.g. in terms of the number of cores or the amount of memory. Such requirements are generally configurable using optional parameters in the coordination language, while the actual mapping on top of adequate worker nodes is left to the implementation of the specific executor.
      
      It is much easier for a scheduling algorithm to work with \textit{homogeneous} resource pools, in which all the nodes have the same characteristics in terms of cores, memory, network and persistence. Nevertheless, in a real scenario, different tasks likely require very diverse amounts of resources, resulting in sub-optimal workloads for homogenous pools. The case of hybrid workflows is even more complicated since the non-uniform data access makes it particularly important to rely on data locality whenever suitable, trying to minimise the need for data transfers among different models.
      
      In general, all container-based WMSs tend to tightly-couple the allocation of a container with the subsequent execution of the task inside it. In this setting, all the available worker nodes are ultimately identified by the amount of computing power they can provide. Requirement $R1$, which states that in \streamflow{}, the unit of deployment should be a complex environment with different containers, introduces an additional level of complexity here. Indeed, it is no longer true that a task can be executed on any worker node equipped with enough hardware, but rather the services exposed by each container can be identified as \textit{capabilities}, and a task can be executed on top of it only if all its \textit{requirements} are satisfied. \streamflow{} straightforwardly manages this requirement-capability association, by identifying each container type with a single service, according to requirement $R2$, and specifying which service is required by each task (through the \texttt{bindings} list described in Sec. \ref{subsec:streamflow-file}).
      Since in \streamflow{} the model life-cycle is managed by an external orchestration library, container-related resources constraints should be specified in the environment description file. Task-related resource constraints, specified in the workflow description, and requirement-capability associations, specified in the \streamflow{} file, are instead directly managed by the \texttt{Scheduler} class when selecting the target resource. Even if only a single target service can be specified for each task, multiple replicas of the same service could exist at the same time and, if the underlying orchestrator provides auto-scaling features, their number could also change in time. It is the responsibility of the \texttt{Scheduler} class to both extract the list of compatible resources for a given task (by calling the \texttt{get\_available\_resources} method of the appropriate \texttt{Connector} instance) and to apply a scheduling policy to find the best target.
      
      \begin{figure}[t]
        \begin{center}
          \includegraphics[scale=0.60]{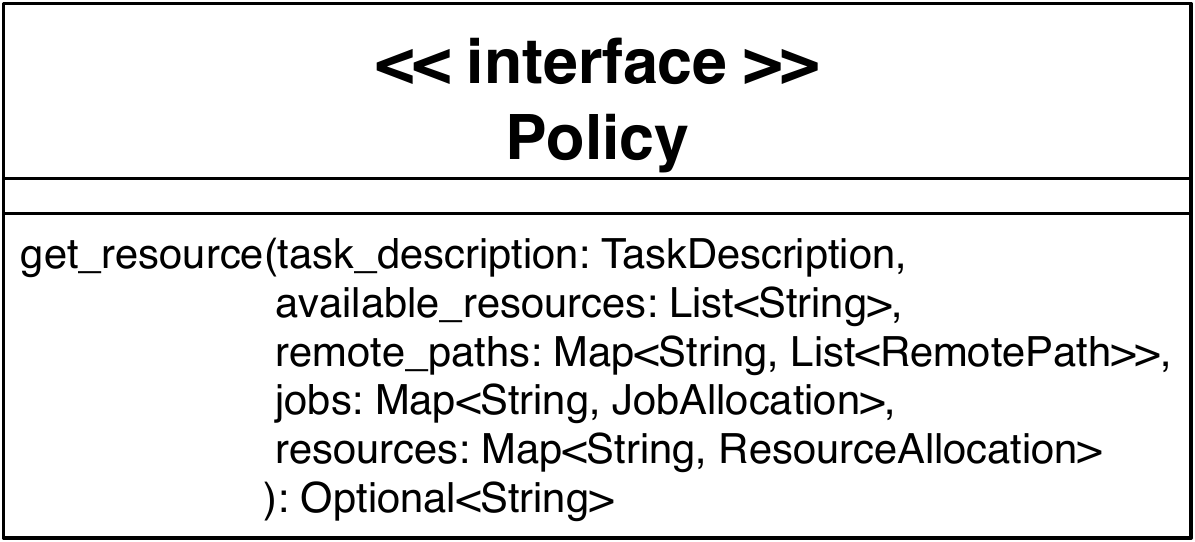}
          \caption{UML class diagram for the \texttt{Policy} interface.} 
          \label{fig:policy-uml-model}
        \end{center} 
      \end{figure}
      Given the very complex nature of the execution environments managed by \streamflow{}, it is improbable that a universally best scheduling strategy actually exists. Indeed, many different factors (e.g. computing power, data locality, load balancing) can affect the overall workflow execution time. For this reason, we decided to implement a \texttt{Policy} interface to allow users to implement their custom strategies. As can be seen from the UML class diagram shown in Fig. \ref{fig:policy-uml-model}, the \texttt{Policy} interface only contains a single method, called \texttt{get\_resource}, with five input arguments:
      \begin{itemize}
        \item The \texttt{task\_description} argument contains a characterisation of the current task in terms of resource requirements and data dependencies.
        \item The \texttt{available\_resources} argument is the list of all the resources which satisfy the requirement-capability association for the current task.
        \item The \texttt{remote\_paths} argument contains, for each file explicitly managed by the WMS, the list of its remote copies. This information can be used by a scheduling policy to take into account data locality in its algorithm.
        \item The last two arguments describe the previously allocated jobs, allowing the implementation of load-balancing features in the scheduling strategy. In particular, the \texttt{JobAllocation} class contains the task description, the resource to which it has been assigned and its status, while the \texttt{ResourceAllocation} class contains the related model and service of an existing resource and the list of jobs assigned to it.
      \end{itemize}
  
      The \streamflow{} \texttt{Scheduler} class processes fireable tasks according to a simple First Come First Served (FCFS) order, without allowing for preemption. Moreover, since each scheduling policy can only process one task at a time, all those strategies that require a global knowledge of the tasks queue (e.g. the various flavours of backfilling or a Shortest Job First approach) cannot currently be implemented. Even if this can result in sub-optimal scheduling solutions in some cases, the proposed approach drastically reduces the implementation complexity, which is an essential aspect for proof-of-concept work.
      
      A very general scheduling policy, serving as a default strategy, comes out-of-the-box with \streamflow{}. When a task becomes fireable, the algorithm iterates over all available resources, starting from those containing at least one of its data dependencies to privilege data locality and trying to reserve the first one which is free (i.e. does not contain jobs in the \texttt{running} status) and satisfies all the constraints. If the search fails, then a \texttt{null} value is returned, and the task is inserted into a waiting queue: a new scheduling attempt will be performed as soon as a \texttt{running} job notifies its termination.
    \end{subsection}
  
    \begin{subsection}{Data transfers} \label{subsec:data-transfers}
      \begin{figure}[t]
        \begin{center}
          \includegraphics[scale=0.60]{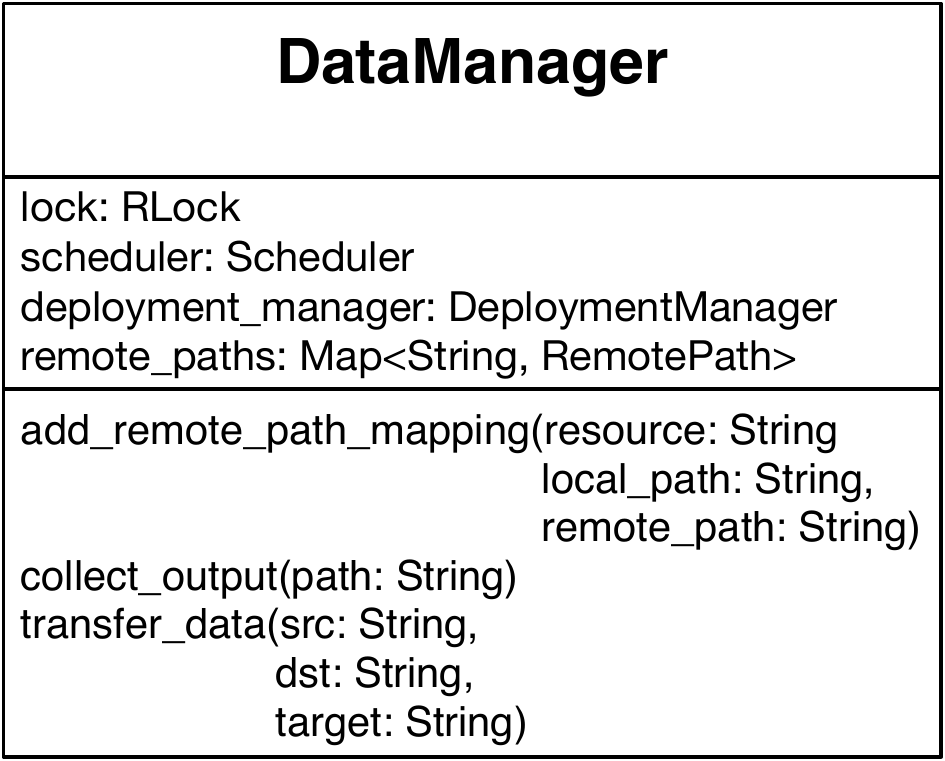}
          \caption{UML class diagram for the \texttt{DataManager} class.} 
          \label{fig:data-manager-uml-model}
        \end{center} 
      \end{figure}
      As pointed out in Sec. \ref{subsec:hybrid-workflows}, hybrid workflow executions make it necessary to waive the comfort brought by a globally shared data space, leaving to the WMS the task of explicitly moving the data whenever required. Since large data transfers are very time-consuming operations, especially for long distances and in the absence of dedicated high-throughput communication networks, the WMS should always select the best communication channel between two endpoints and avoid all unnecessary data movements. The \streamflow{} framework has been designed in order to meet requirements $R3$ and $R4$, which represent two fundamental steps in this direction. In particular, a dedicated \texttt{DataManager} class, whose UML diagram is in Fig. \ref{fig:data-manager-uml-model}, has been developed with the precise goals of keeping track of the remote locations of each data dependency and performing data transfers between successive steps.
      
      Whenever a task terminates in \texttt{completed} status, it is in charge of atomically updating the \texttt{remote\_paths} structure with the remote position of all its output files and folders by calling the \texttt{add\_remote\_path\_mapping} method. The same structure is also used by the \texttt{transfer\_data} method, which is called every time a task needs a file or a folder from one of its predecessors, to verify if a data transfer is needed or not. In particular, transfers can always be avoided when both tasks run on the same resource, but this can also happen when two resources share a data space (e.g. a persistent volume) or if a task explicitly performs a data transfer before completing.
  
      If the destination path does not exist, then a data movement is unavoidable. If the source and the target resources belong to distinct models, then \streamflow{} adopts the baseline strategy mentioned in requirement $R3$, performing the first transfer from the source resource to the management node and a second copy to the target resource. Instead, if the two resources belong to the same model, the transfer is directly performed by the \texttt{copy} method of the corresponding \texttt{Connector} implementation. In the latter case, some optimisations are possible. For example, since all Occam nodes share the \texttt{/archive} and \texttt{/scratch} portions of the file-system, only a local copy on the target resource is required to transfer a data dependency which resides in one of such folders.
  
      Finally, the \texttt{collect\_output} method performs a data transfer from a remote resource to the local management node. This method is always called before a remote resource is undeployed in order to retrieve the final output of the workflow model. Moreover, when a task must be performed locally but requires some remote input data, this method is called before starting its execution.
    \end{subsection}
    
    \begin{subsection}{Task-container mapping patterns} \label{subsec:mapping-patterns}
        Since models are not redeployed after each task execution, when multiple tasks are bound to the same service \streamflow{} implements by default an MTSC pattern. This design choice is the standard one adopted by CWL when tasks are executed on the local machine and is consistent with $R4$, which tries to minimise data transfers to achieve better performances. Nevertheless, it gives up for the clean and consistent execution environment commonly provided by containers, which can be problematic if previous task executions can have unexpected effects on the next ones. In order to force an STSC pattern, a \texttt{recycle} directive can be added to a binding entry in the \streamflow{} file. This induces a redeployment of the involved service before the task execution. In such a case, \streamflow{} will automatically handle all required data transfers, ensuring that at least one copy of each task output is stored in a persistent location before deleting the container.
        
        The unique parallel execution pattern natively included in CWL standard is the \texttt{scatter}, in which a list of input data is partitioned among multiple, identical tasks that can be executed in parallel by multiple nodes. More complex interactions among tasks (e.g. an MPI application) must be directly handled in the code, and it is up to the user to ensure that the correct amount of worker nodes is up and running before the task execution. Conversely, \streamflow{} offers explicit support for STMC mapping: in the \streamflow{} file, a single step can be bound to multiple resources through the \texttt{replicas} directive (which defaults to 1). In case of multiple replicas, \streamflow{} initialises two additional environment variables: a \texttt{STREAMFLOW\_RANK} variable, containing a unique rank for each job, and a \texttt{STREAMFLOW\_HOSTS} variable, containing the comma-separated list of nodes (i.e. hostnames) allocated for the task. These variables can be used inside the task script to guide the execution on each node, e.g. to implement a master-worker pattern or to execute an \texttt{mpirun} command on the node with rank 0.
        
        The case of MTMC pattern is a bit trickier. Indeed, while \streamflow{} seamlessly supports co-allocation of different tasks on different services, the CWL standard is not able to explicitly describe such property. Although the simultaneous allocation of apparently independent steps just happens in many situations, it is evident that a formal and generic way to express tasks co-allocation directly in the workflow model is a mandatory requirement to entirely support this feature. Find the best way to improve MTMC pattern support is an essential milestone in \streamflow{}'s future development.
    \end{subsection}
  \end{section}
  
  \begin{section}{Single-Cell application use-case}
  \label{sec:usecase}
    As extensively described in Sec. \ref{subsec:scientific-workflows}, scientific applications are an ideal target for workflow modelling. To demonstrate \streamflow{}'s ability to satisfy requirements at both user-level, in terms of supporting easy application modelling, and infrastructure-level, offering flexibility in the choice of deployment targets, we selected a novel pipeline in the field of Bioinformatics, which is the discipline supporting molecular biology and biomedicine in the analysis of data. Bioinformatics, together with astronomy, is one of the first scientific fields to deal with Big Data. Indeed, biological datasets are massive, heterogeneous, and grow very fast, making this discipline hungry for computational power and storage capabilities.
    
    Moreover, these data should be analysed by Bioinformatics researchers, with a mixed background in biology, medicine and computer science. This scenario requires the development of hybrid computational systems by HPC experts, implying  a careful selection of the target infrastructure according to the different applications, desired performance, but also cost requirements. 
    
    From this perspective, in this section, we want to show how \streamflow{} features can be used to implement complex analysis workflows in an extremely portable way, which allows users to find the best deployment option for each step in heterogeneous, hybrid HPC/cloud infrastructures without modifying neither the code nor the workflow description itself.

    \begin{subsection}{Single-cell sequencing} \label{subsec:single-cell-sequencing}
        More specifically, the Bioinformatic application we selected for our tests is a pipeline for single-cell sequencing data analysis. Generally stated, sequencing is the process of determining the order of the four bases (adenine, guanine, cytosine, and thymine/uracil) of a nucleic acid molecule, which can be DNA or RNA. The first nucleic sequences were obtained in the early 1970s by academic researchers using laborious methods based on two-dimensional chromatography. Following the development of fluorescence-based sequencing methods, sequencing has become more accessible and orders of magnitude faster, allowing the first draft of the human genome. During the last decade, massive high-throughput sequencing methods have revolutionised the entire field of molecular biology, both considering DNA sequencing and RNA sequencing, accelerating medical research and discovery, since samples from patients can now be sequenced routinely.
    
        DNA sequencing is usually performed to describe the genomic differences between two samples, which impact on their phenotypes, such as having different eyes' colour or susceptibility to a specific disease. On the other hand, RNA sequencing (RNA-seq) is performed to understand what is going on inside the cell, which genes are actually transcribed and active, because, for example, a liver must implement different biological processes in comparison to a spleen (although they share precisely the same genome). The idea is that we can compare DNA to a program stored on a disk and RNA as the same program loaded into RAM.
    
        The opportunity to study the transcriptome of cells (cultured cells, cells from mouse models or cells from human samples, such as blood) using RNA-seq has fuelled many crucial discoveries in biology and biomedicine, being now a routine method in clinical research. However, RNA-seq is typically performed in "bulk", which means to sequence the RNA of all the cells in a sample (thousands or millions of cells), and the data represent an average gene expression pattern across a population of cells. This might obscure biologically relevant differences between cells, such as tumour clones that are resilient to chemotherapy.
        
        Single-cell RNA-seq (scRNA-seq) represents an approach to overcome this problem. The idea is to isolate single cells through microfluidic approaches, capturing their transcripts through emulsion droplets loaded with chemical reagents, and generating sequencing libraries in which the transcripts are tagged (through a nucleotidic barcode) to track their cell of origin. One of the most popular platforms for single-cell analysis is marketed by 10X Genomics, which is capable of analysing from 500 to 20,000 cells in each run. Then, combined with massive high-throughput sequencing producing billions of reads, scRNA-seq allows the assessment of fundamental biological properties of cells populations and biological systems at unprecedented resolution. 
    
        The problem with this technique is the noise that is exaggerated by the need for very high amplification from the small amounts of RNA found in each cell. Denoising these data and estimating the adequate amount of sequencing reads covering each gene in the cell is of critical importance to define a reliable RNA \textit{count matrix}, the fundamental data structure for this kind of analysis which represents for each cell and each gene how many transcripts have been captured. This is a quite complex bioinformatic pipeline that requires many different statistics and repeating the procedures many times to identify the right thresholds for the sample in analysis. In this context, the processing power and the automatic management of the pipeline are of critical importance, since analysing each cell in a population requires from hundreds-of-thousands to millions of comparisons to be processed in a high throughput manner.
    \end{subsection}

    \begin{subsection}{Application pipeline} \label{subsec:application-pipeline}
      Once the noise caused by the experimental amplification of the RNA has been controlled, and the count matrix has been built, the key idea is to implement techniques aimed at reducing the data dimensionality in order to cluster cells with a similar expression profile. Therefore, a typical pipeline for single-cell transcriptomic data analysis can be broadly divided into two main parts: the creation of the count matrix and its statistical analysis.
      
      The first step is the creation of the RNA count matrix, and it must be performed according to the adopted single-cell experimental technology and the used sequencing approach. For example, considering a typical 10x genomics experiment followed by an Illumina Novaseq sequencing, the first part of the pipeline will be performed using a tool called CellRanger \cite{Zheng:2017}. In particular, this part of the analysis will consist in two steps: the creation of the fastq files (the raw sequences of the four bases, called reads) from the flowcell provided in output by the sequencer and the alignment of the reads against the reference genome.
  
      The fastq creation is performed by looking at the images generated by the sequencer cycle after cycle into the flowcell on which the sequences have been hybridised. From the computational point of view, the algorithm looks at the images and calls the bases for each position. It also provides, for each base in each read, a quality score according to the accuracy by which the base has been called. 
  
      The second step performed by CellRanger is the creation of the count matrix itself, a process that requires two distinct procedures. First, sequences that have been generated in the previous step are aligned against the reference genome using STAR, which is the most popular aligner currently available for transcriptomic analysis. These alignments are then processed according to the genome annotation, in order to recapitulate for each gene how many reads have been captured. 
      
      Once the count matrix has been computed, a quantitative analysis of the results is usually performed. The aim is clustering cells having similar transcriptomic profiles and characterising them according to some reference databases. This can be performed using ad-hoc developed software in Python or R, the latter being probably the most popular at the moment. In the context of this pipeline, we used two main R packages for the analysis of the count matrix: Seurat \cite{Butler:2018,Stuart:2019} for normalisation, dimensionality reduction and clustering of cells, and SingleR \cite{Aran:2019} for labelling the clusters, that is identifying the cell type, according to public databases of single-cell data annotation.   
  
      In particular, Seurat is used to loading data into the R environment and to filter outliers for specific statistics, such as the number of unique transcripts or the presence of mitochondrial transcripts, which correlate with the vitality of the cells. Data are then normalised, taking into account the different coverage of the different cells, and the most variable genes are identified. These genes are used to perform a dimensionality reduction through the computation of principal component analysis. Cells are then clustered using the Louvain algorithm, which has been specifically designed for detecting communities in networks. At last, marker genes are identified for each cluster by comparing the expression profile of the cells inside the cluster with all the other cells. 
     
      Once clusters have been identified, the pipeline uses SingleR to characterise each cell trying to identify its type (such as Blood Cell, Bone Cell, and Stem Cell) in an unbiased way. SingleR leverages reference transcriptomic datasets of pure cell types to infer the identity of every single cell independently. In particular, SingleR starts by calculating a Spearman coefficient for each cell in the single-cell experiment with the reference data set, using only variable genes, thus increasing the ability to distinguish closely related cell types. This process is performed iteratively, using only the top cell types from the previous step and only the variable genes among these remaining cell types, until a precise cell type can be assigned to the analysed cell. 
      
      As a test case, in this work, we used a published dataset \cite{Schiroli:2019} concerning Gene Editing in Hematopoietic Stem Cell. In particular, this dataset was produced to compare the efficiency of different gene-editing approaches and, for this reason, the whole experiment is composed of 6 different single-cell samples sequenced independently. This complex experimental design resulted in a particularly challenging and time-consuming dataset, making a flexible, automated and scalable workflow management systems particularly desirable.
      
    \end{subsection}
  
    \begin{subsection}{\streamflow{} implementation}
        \begin{figure}[t]
            \begin{center}
              \includegraphics[width=.6\textwidth]{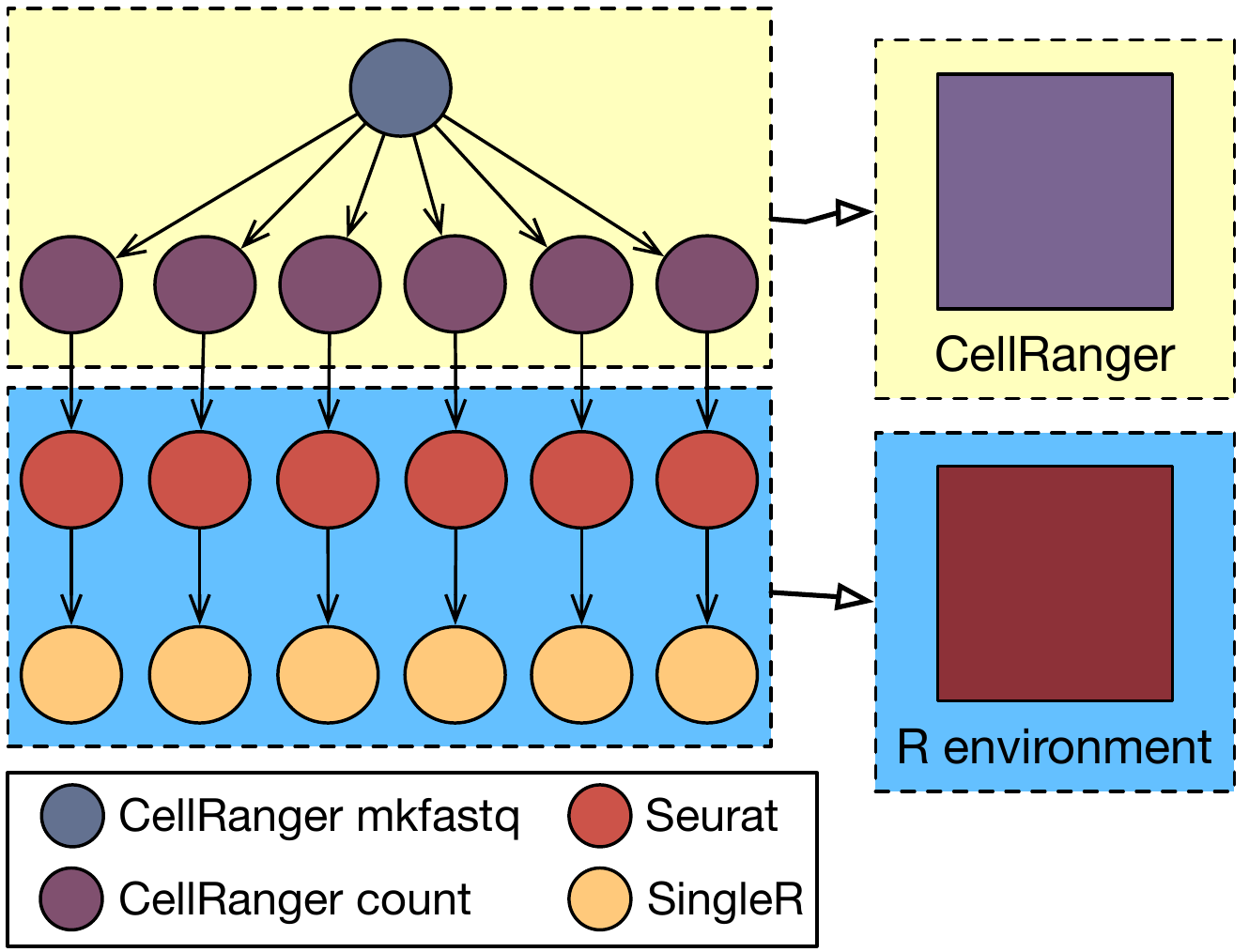}
              \caption{Dependency graph and model bindings for the single-cell workflow. In this case, the first step creates six different sequences, which can then be processed independently of each other for the remaining three steps.}
              \label{fig:single-cell-wf-model}
            \end{center} 
      \end{figure}
      In general, the design of a \streamflow{} application can be split into three high-level steps:
      \begin{itemize}
        \item The design of the workflow dependency graph, using a coordination language of choice among those supported by the framework. For now, as discussed in Sec. \ref{subsec:wms-integration-layer}, CWL is the only possible choice, so we obviously opted for it.
        \item The design of the execution environment, in terms of one or more models containing one or more services each. Here we decided to experiment two different combinations of Occam and Helm environments, as better detailed below in this section.
        \item The creation of a \streamflow{} file, as described in Sec. \ref{subsec:streamflow-file}, in order to wrap things together.
      \end{itemize}
    
      Fig. \ref{fig:single-cell-wf-model} provides a graphical representation of the whole \streamflow{} model for a single-cell pipeline of the kind described in Sec. \ref{subsec:application-pipeline}. In this case, the workflow dependency graph is a simple DAG with four different kinds of tasks. In terms of workflow patterns \cite{vanderAalst:2003}, it can be represented as an initial parallel split, with a fan-out equal to the number of sequences produced by the first task (six in this case), followed by as many independent sequence blocks of three tasks each. In CWL, this can be easily implemented using the \texttt{scatter} directive. It is also worth noting that, since none of the tasks can be executed in a distributed fashion, the maximum number of nodes from which the workflow execution can take some benefit is equal to the fan-out of the initial parallel split.
      
      Since CellRanger executes the first two types of tasks and the last two tasks require two main R packages (i.e. Seurat and SingleR) plus all the related dependencies, we decided to implement two distinct container images. Partitioning the tasks with respect to their target container, we obtain two disjoint subsets, each of which can execute concurrently on a maximum of six nodes. Therefore, if enough hardware resources are available, the best strategy would be to allocate six replicas of each image, implementing an MTSC mapping as described in Sec. \ref{subsec:mapping-patterns}. Nevertheless, since the containers initialisation time is negligible with respect to the time required for the completion of tasks themselves and outputs are always stored in a persistent location, this ends up being practically equivalent to an STSC pattern. Given that, it should be clear that requirements $R1$ and $R2$ of Sec. \ref{subsec:multi-container-environments} do not bring additional concrete value to this workflow.
  
      Conversely, the hybrid workflow execution enabled by requirements $R3$ and $R4$ of Sec. \ref{subsec:hybrid-workflows} can be beneficial, for example, to perform a data preprocessing phase on a dedicated HPC structure before moving data to the cloud to complete the remaining steps. Indeed, in the examined case, the total size of the initial data is almost 60GB, but modern sequencing machines can achieve 10 billion of sequences per flowcell, corresponding to about 3TB of data. Moreover, the \texttt{cellranger count} command requires a quite high amount of resources to be performed: the official documentation reports 8 cores and 32GB of memory as minimum requirements, but a significant speedup can be appreciated until up to 32 cores and 128GB of memory.
  
      If hybrid workflows were not supported, the best strategy would be to execute the entire set of tasks on top of six HPC nodes, in order to take full advantage of the available grade of parallelism while avoiding data transfers. Moreover, when using total wall clock time as the only evaluation metric, this one keeps being the best solution also when compared with hybrid alternatives. Therefore, it is worth to use this setting as a baseline in order to evaluate the significance of performance loss when switching to a mixed HPC/cloud configuration.
      
      We reserved six Light nodes on the Occam facility, each of which having 2x Intel Xeon E5-2680 v3 (12 core each, 2.5GHz) CPUs and 128GB (8x16, 2133MHz) of memory, and prepared a model which allocates each node to both a CellRanger and an R environment containers. As mentioned in Sec. \ref{subsec:data-transfers}, all Occam nodes share the \texttt{/archive} folder, mounted as an NFS export, and the \texttt{/scratch} folder, with a LUSTRE parallel file-system. We copied initial data on the \texttt{/archive} file-system and configured \streamflow{} to use a folder on the \texttt{/scratch} hierarchy as its output folder. In this way, data could be accessed by dependent tasks without the need for explicit transfers. Then we ran the \streamflow{} application inside a container launched on an additional Occam node.
      
      \begin{figure}[t]
        \begin{center}
          \includegraphics[width=.9\textwidth]{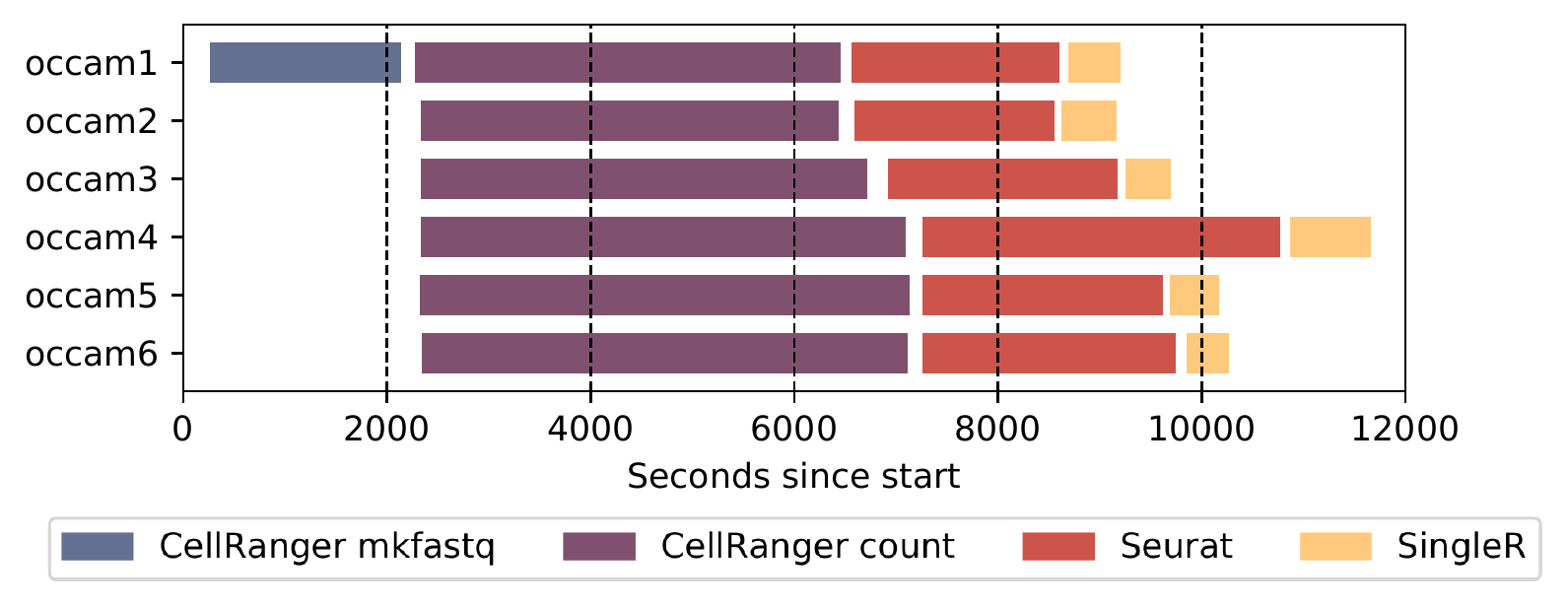}
          \caption{Execution timeline for the \streamflow{} single-cell application on six Occam nodes, each allocated to both a CellRanger and an R environment containers.} 
          \label{fig:occam-timeline}
        \end{center} 
      \end{figure}    
      The timeline for this execution is reported in Fig. \ref{fig:occam-timeline}. The whole duration is about three hours and a quarter, dominated by the CellRanger count and the Seurat commands. White space between subsequent bars represents the time needed by \streamflow{} itself to perform some internal tasks before launching a new command, including copying the input data on a staging folder (as mentioned in Sec. \ref{subsec:data-transfers}). Nevertheless, the time taken to perform each of these operations is negligible with respect to the time needed to complete tasks themselves.
  
      In a real scenario, it would be probably better to dedicate the HPC structure to the completion of the first tasks, while executing the rest of the workflow directly on a cloud environment. Indeed, the output data of the last task must often be stored into a database or visualised in a web application, and the cloud is undoubtedly the most natural place to host such kind of services. By observing intermediate data in the workflow model, it is possible to notice that output data of the second task have a total size of about 15-30MB, while the third task produces output data for more than 200MB. Given that, in order to minimise the overhead introduced by a data transfer, the best strategy would be to execute the first two tasks on an HPC facility and the remaining two in a cloud infrastructure.
      
      We configured a virtualised Kubernetes cluster on top of the GARR\footnote{\url{https://garr.it/it/}} cloud, based on OpenStack\footnote{\url{https://www.openstack.org/}}, containing six worker nodes with 4 virtual CPUs and 8GB of memory each. Then we prepared two different models:
      \begin{itemize}
        \item A first model with six Occam nodes, with an instance of the CellRanger container allocated on each of them
        \item A second model with six Kubernetes Pods, each with an instance of the R environment container and a \texttt{podAntiAffinity} parameter to ensure that each Pod is allocated on a different worker node whenever possible.
      \end{itemize}
      It is worth noting that there is no need to modify the CWL description of the workflow to run it on the new environment: changes only involve model descriptions and the \texttt{streamflow.yml} file.
      
      On Kubernetes, the \streamflow{} output folder of each container has been mapped to a persistent volume managed by Cinder, the OpenStack's block storage service, configured with a \texttt{readWriteOnce} access mode. It means that no shared data space exists between different worker nodes. Nevertheless, the scheduling policy described in Sec. \ref{subsec:task-scheduling} makes it so that each SingleR task is executed by the node where its required input data already reside, removing the need for additional data transfers. Given that, since we kept running \streamflow{} application inside an Occam node, the only unavoidable data movement is from Occam to Kubernetes, between the second and the third tasks.
  
      \begin{figure}[t]
        \begin{center}
          \includegraphics[width=.9\textwidth]{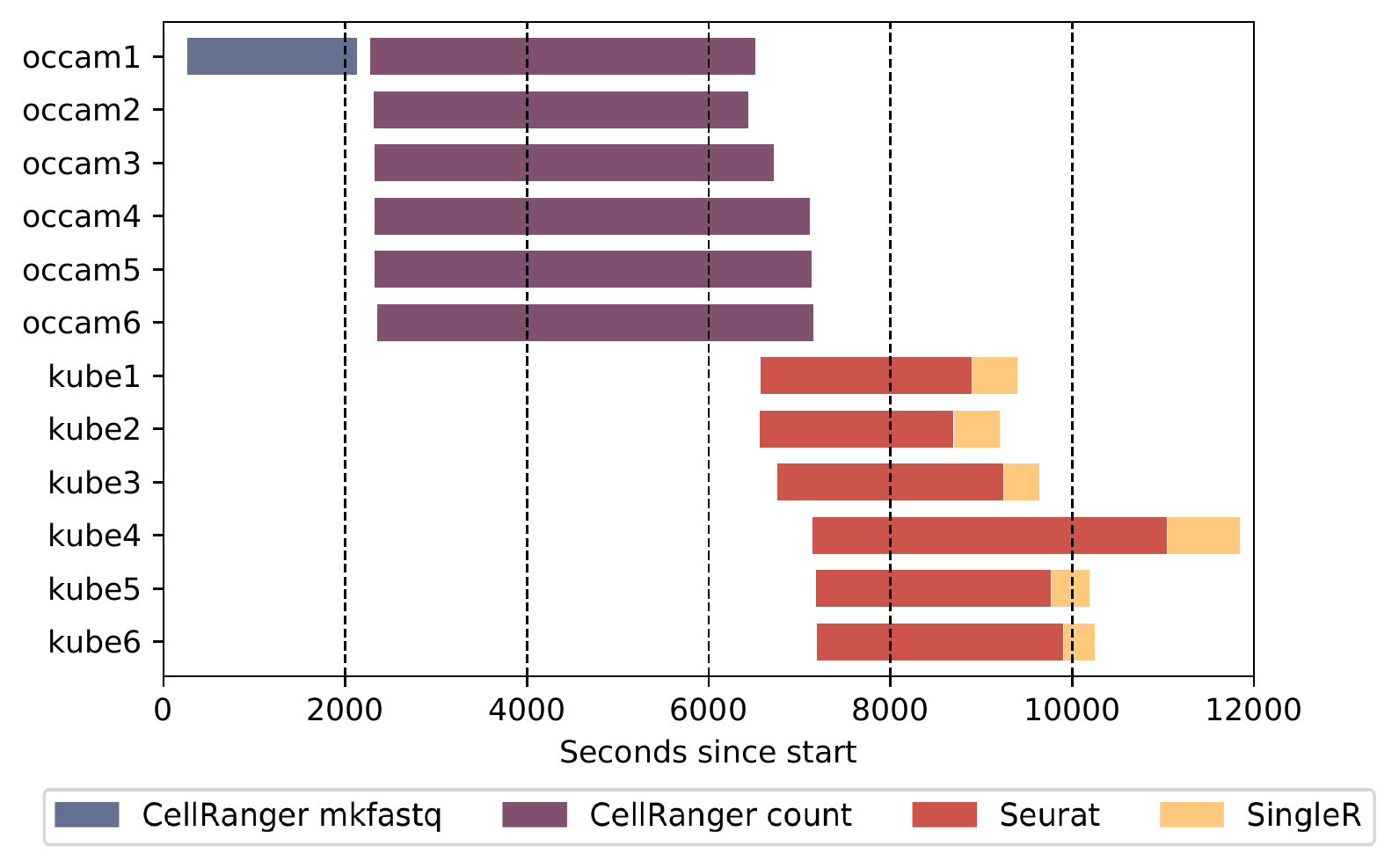}
          \caption{Execution timeline for the \streamflow{} single-cell application in a hybrid configuration, with six Occam nodes allocated to CellRanger as many replicas and six Kubernetes worker nodes allocated to as many R environment containers.} 
          \label{fig:occam-helm-timeline}
        \end{center} 
      \end{figure}
      The timeline for this second run is reported in Fig. \ref{fig:occam-helm-timeline}. The first important thing that can be observed is how the whole duration of this hybrid execution is comparable with the previous full-HPC configuration. This is mainly due to the combination of two factors. Firstly, the time needed to transfer data from the Occam facility to the GARR cloud is negligible when compared with the time needed to complete the tasks themselves. Moreover, the Seurat task seems not to benefit so much from additional computing power, making it quite useless to commit HPC machines for its execution. In a situation like this, it is pretty clear that the \streamflow{} approach can be beneficial to obtain a more efficient resource allocation without significant performance drops.
  
    \end{subsection}
  \end{section}
  
  \begin{section}{Conclusion and further development}
  \label{sec:conclusion}
    The recent explosion in popularity faced by lightweight containerisation technologies also invested the scientific workflows' ecosystem, with undoubted gains in portability and reproducibility. During the very last years, a significant number of WMSs started to include container-based workflow executions among their features, while new container-native alternatives began to appear. Nevertheless, some common simplifications in the design process can prevent a WMS to exploit the potential of containerisation technologies fully.
  
    This work aims at exploring the potential benefits deriving from waiving two common properties of existing WMSs. Firstly, a one-to-one task-container mapping prevents the execution of tasks in multi-container environments and makes it unnecessarily difficult to support concurrent executions of communicating tasks. Moreover, the requirement for a single shared data space represents an obvious obstacle for hybrid workflow executions, which could instead highly benefit from containers' portability properties.
  
    The \streamflow{} framework has been developed as a proof-of-concept WMS which explicitly drops these constraints by design. In \streamflow{}, the unit of deployment is a complex multi-container environment, directly managed by an underlying orchestration technology. Moreover, each container can exchange files with every other, with the only constraint for the WMS management node to be able to reach the whole execution environment. This second feature has been used to run a bioinformatics workflow on top of a hybrid HPC/cloud environment without significant performance losses, therefore showing the potential benefits introduced by the proposed approach in terms of more efficient resource usage.
  
    The next crucial step now is to investigate benefits brought by multi-container deployment units in scientific applications. Potential forthcoming candidates for experimentation are all those applications which require distributed execution, as MPI-based simulations or distributed deep learning frameworks. In case of positive feedback, some further developments will be necessary to evolve \streamflow{} in a mature product, as the support for more coordination languages and orchestration libraries. Moreover, as previously mentioned, a robust abstraction for inter-container communication channels would significantly reduce performance losses introduced by large data transfers.
  \end{section}

\section*{Acknowledgement}
\begin{wrapfigure}{l}{0.15\textwidth}
\begin{center}
\vspace{-22pt}
\includegraphics[width=0.15\textwidth]{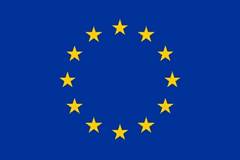}
\vspace{-24pt}
\end{center}
\end{wrapfigure}
  This article describes work undertaken in the context of the DeepHealth project\footnote{\url{https://deephealth-project.eu/}}, \emph{``Deep-Learning and HPC to Boost Biomedical Applications for Health''} which has received funding from the European Union's Horizon 2020 research and innovation programme  under grant agreement No. 825111 \cite{8787438}. This work has been partially supported by the HPC4AI project\footnote{\url{http://www.hpc4ai.it}} \cite{18:hpc4ai_acm_CF}.




%


\end{document}